\documentclass[fleqn,usenatbib]{mnras}

\usepackage{newtxtext,newtxmath}
\usepackage[T1]{fontenc}
\usepackage{ae,aecompl}

\usepackage{graphicx}	% Including figure files
\usepackage{amsmath}	% Advanced maths commands
\usepackage{amssymb}	% Extra maths symbols
\usepackage{subcaption}
\title[Galactic Centre Generalised Bondi]{A Generalised Bondi Accretion Model for the Galactic Centre}

\author[A. Yalinewich et al.]{
A. Yalinewich,$^{1}$\thanks{E-mail: almog.yalin@gmail.com}
R. Sari,$^{2}$
A. Generozov $^{3}$
N. C. Stone $^{3}$
and B. D. Metzger$^{3}$
\\
$^{1}$Canadian Institute for Theoretical Astrophysics, 60 St. George St., Toronto, ON M5S 3H8, Canada\\
$^{2}$Racah Institute of Physics, the Hebrew University, 91904, Jerusalem, Israel\\
$^{3}$Columbia Astrophysics Laboratory, Columbia University, 550 W. 120th Street, New York, NY, 10027, USA
}

\date{Accepted XXX. Received YYY; in original form ZZZ}

\pubyear{2018}

\begin{document}
\label{firstpage}
\pagerange{\pageref{firstpage}--\pageref{lastpage}}
\maketitle

% Abstract of the paper
\begin{abstract}
We develop an analytic, steady-state model for the gas environment in quiescent galactic nuclei. We assume that the mass is constantly supplied by a spherically symmetric distribution of wind emitting stars, and that gravity is solely due to a central supermassive black hole. We show that at some finite radius, where the Keplerian velocity is comparable to the wind velocity, the bulk velocity vanishes. Matter generated below that radius will be accreted onto the black hole, while matter outside it will escape the system. Under certain conditions, the flow may become supersonic at both domains. We obtain radial profiles of the hydrodynamic variables and verify them using a time-dependent hydrodynamic simulation. We delineate the conditions under which radiative cooling can be neglected, and predict the luminosity and spectrum of the free-free X-ray emission from such a system.  We discuss applications of our solution to our own Galactic Centre and other quiescent galactic nuclei.
\end{abstract}

% Select between one and six entries from the list of approved keywords.
% Don't make up new ones.

%%%%%%%%%%%%%%%%%%%%%%%%%%%%%%%%%%%%%%%%%%%%%%%%%%

%%%%%%%%%%%%%%%%% BODY OF PAPER %%%%%%%%%%%%%%%%%%

\section{\label{sec:introduction} Introduction}

In this paper we develop an analytic, adiabatic, steady state, hydrodynamic model to describe the environment shaped by constant effluence from stellar winds and the gravity of a central point mass, such as a supermassive black hole. Our main application is to galactic nuclei like our own galactic centre, which has been observed over the past few decades at high spatial resolution \citep{genzel_et_al_2003}.  In particular, observations by the {\it Chandra X-ray Observatory} enabled measurements of the density and temperature of the hot gas located less than a parsec from the central black hole, SgrA* \citep{baganoff_et_al_2003}.  

Similar observations have been used to probe the properties of gas in the nuclei of other galaxies.  Though most galactic nuclei are dim and inactive like SgrA* \citep{ho_2009}, they are occasionally ``lit up'' by tidal disruption events \citep{rees_1988}.  The latter occur when a star in the nuclear cluster is placed onto an orbit that brings it too close to the supermassive black hole, such that tidal forces from the black hole overcome the self-gravity of the star and tear it apart, producing a transient accretion event.  Some of these events produce radio synchrotron emission (e.g.~\cite{zauderer_et_al_2011,berger_et_al_2013}), which was predicted to result from a shock wave driven into the gas by a relativistic jet or outflow powered by the accretion flow \citep{giannios_metzger_2011}.   Under certain assumptions, it is possible to infer the radial profile of the gas density in the galactic centre using the measured flux and spectral evolution of this radio emission.

The present study complements previous analytical and numerical studies of wind-shaped environments.  One of the earliest works was by Parker \citep{parker_1958}, who considered the evolution of a fluid parcel as it leaves the stellar surface and is accelerated outwards by pressure forces, becoming supersonic after passing through the critical point.  The Parker wind solution is the reverse case of Bondi accretion \citep{bondi_1952}, with matter flowing outward rather than inward. \cite{chevalier_clegg_1985} considered a spherically symmetric, non-gravitating system with constant mass injection per unit volume inside a certain radius, and no mass injection outside that sphere. They found an analytic, albeit implicit, solution to the steady state hydrodynamic equations in this case for the spatial profiles of the density, velocity and temperature. 

\cite{quataert_2004} numerically integrated the steady-state hydrodynamic equations in order to model the diffuse {\it Chandra} X-ray measurements around SgrA*.  His model accounted for the gravity of the black hole and a source term of gaseous mass injection confined between two radii meant to mimic the input from the observed cluster of massive stars. This model was later refined by \cite{Shcherbakov_Baganoff_2010}, who represented each wind emitting star as a thin spherical wind emitting shell with a radius equal to the galactocentric distance of the star, and by \cite{2018arXiv180500474R}, who performed a three dimensional simulation with wind emitting stars and point sources.  Similar models have also been applied to the study of globular clusters \citep{2013arXiv1310.8301N} and nuclear star clusters \citep{de_colle_et_al_2012, shcherbakov_et_al_2014, generozov_et_al_2015}.  In general, the solution will depend on the specific heating (and cooling) rate of gas injected by stellar winds, which in turn depends on the age of the stellar population and on the radial density profile of the stars.    

This paper is organised as follows.  In section \ref{sec:analytic} we present the mathematical formulation of the problem and discuss some limiting cases. In section \ref{sec:numerical_sim} we verify our analytic results using a time dependent hydrodynamic simulation. In section \ref{sec:radiation} we discuss the effects of radiation. In section \ref{sec:astro_app} we discuss the astrophysical application of the model developed. Finally, in section \ref{sec:summary} we summarise the work.

\section{Analytic Estimates} \label{sec:analytic}

\subsection{Governing Equations}

We are interested in the steady state of a system where mass and energy are continuously injected, and matter either flows out to infinity or is accreted onto gravitating mass at the centre.  The radial location separating inflow on small scales and outflow on large scales, where the velocity passes through zero, is known as the {\it stagnation radius,} $r_{\rm st}$.  The system is governed by the following hydrodynamic equations
\begin{subequations}\label{eq:model}
\begin{align}
  &\frac{\partial \rho}{\partial t}+\frac{1}{r^2}\frac{\partial}{\partial r}\left(\rho r^2 u\right)=q \label{eq:mass}\\
  &\rho \frac{\partial u}{\partial t} + \rho u\frac{\partial
    u}{\partial r}=-\frac{\partial p}{\partial r}-\rho
  \frac{{\rm d} \Phi}{{\rm d} r} -q u \label{eq:mom}\\
  &\frac{\partial \epsilon}{\partial t}+ \frac{1}{r^2}
  \frac{\partial}{\partial r} \left[\rho r^2 u \left(\epsilon
    +\frac{p}{\rho}\right)\right]=\frac{1}{2} q v_w^2+q\Phi \nonumber\\
  &\epsilon \equiv \frac{1}{2} v^2 + \Phi + \frac{p}{\rho (\gamma-1)}
\label{eq:energy}
\end{align}
\end{subequations}
which account for conservation of mass, radial momentum, and internal energy, respectively.  Here $\rho$, $u$, $p$, $\epsilon$, $\gamma$ are the gas density, velocity, pressure, specific internal energy, and adiabatic index, respectively.   The gravitational potential,
\begin{equation}
\Phi = -\frac{G M}{r}, \, 
\end{equation}
is taken to be that of a point mass, where $G$ is the gravitational constant and $M$ is the mass of the central black hole.  We ignore additional contributions to the gravitational potential from stars or dark matter.  

The stellar wind velocity $v_w$ determines the specific rate of local energy and momentum injection.  We take $v_w$ to be constant with radius.  In reality, close to the black hole the stellar orbital velocity will exceed the wind velocity and $v_{\rm w}$ should increase with the Keplerian velocity $\propto r^{-1/2}$; however, we ignore this correction in this work, because in most astrophysical settings of interest, mass and energy injection are dominated by larger radii. 

We parametrise the mass injection rate per unit volume, $q(r)$, as a power-law in radius,
\begin{equation}
q = D r^{-\eta}
\label{eq:eta}
\end{equation}
where $D$ and $\eta$ are constants.  

Our approach neglects the granularity of mass injection \citep{li_bryan_et_al_2017,rockefeller_et_al_2005}.  Rather than discrete point sources, we assume that matter and energy are injected following a spatially smooth profile.  This is a reasonable approximation for radial distances from the black hole which greatly exceed the average distance between wind-emitting stars.  Although this assumption obviously breaks down at small radii, in most cases of interest mass and energy injection are dominated by large radii, where many stars contribute and granularity is modest.

We also ignore the effects of angular momentum.  A non-zero net angular momentum will have little effect at large radii, but will cause the inflowing gas to stall at a small radii due to the centrifugal barrier.  The behaviour of the accretion flow interior to this point has been explored in many other works (e.g.~\citep{roberts_jiang_et_al_2017, inayoshi_ostriker_et_al_2018}), but here we assume it does not affect the gas flow on larger scales.  Kinetic and radiative feedback from the inner accretion disk are neglected, except insofar as its effects can be captured by the value of $v_{\rm w}$.

\subsection{\label{sec:dimles_red} Reduction to Dimensionless Variables}
The dimensional parameters of the problem are:
\begin{enumerate}
\item $GM$ -- Mass of the central black hole
\item $v_w$ -- Stellar wind velocity (specific heating rate)
\item $D$ -- gaseous mass source term normalization,
\end{enumerate}
%We note that $M$ never appears on its own in the hydrodynamic equations, but always multiplied by $G$, so technically the parameter should be $G M$. 
from which we can define dimensionless variables:
\begin{equation}
\tilde{r} \equiv \frac{r}{R_b}, \, \tilde{\rho} \equiv \frac{\rho}{\rho_b}, \, \tilde{u} \equiv \frac{u}{v_w}, \, \tilde{c}_s \equiv \frac{c_s}{v_w} \, , \, \tilde{p} \equiv \frac{p}{\rho_b v_w^2} \, ,
\end{equation}
where the length scale of the problem is the Bondi radius,
\begin{equation}
R_b \equiv \frac{G M}{v_w^2} \label{eq:bondi_radius},
\end{equation}
and the density scale is defined as
\begin{equation}
\rho_b \equiv D R_b^{1-\eta}/v_w = D \frac{G^{1-\eta} M^{1-\eta}}{v_w^{ 3-2 \eta}}.
\end{equation}
We also define a dimensionless stagnation radius, $\tilde{r}_{st} \equiv r_{\rm st} / R_b$.

 Written in terms of the dimensionless variables, equations \ref{eq:mass}$-$\ref{eq:energy} become
\begin{equation}
\frac{1}{\tilde{r}^2} \frac{d}{d \tilde{r}} \left(\tilde{\rho} \tilde{u} \tilde{r}^2 \right) = \tilde{q} = \tilde{r}^{-\eta},
\label{eq:dimlessmass}
\end{equation}
\begin{equation}
\tilde{\rho} \tilde{u} \frac{d \tilde{u}}{d \tilde{r}} + \frac{d \tilde{p}}{d \tilde{r}} = -\tilde{q} \tilde{u} - \frac{\tilde{\rho}}{\tilde{r}^2},
\label{eq:dimlessmom}
\end{equation}
\begin{equation}
\frac{1}{\tilde{r}^2} \frac{d}{d \tilde{r}}\left[ \tilde{r}^2 \tilde{u} \left(\frac{1}{2} \tilde{\rho} \tilde{u}^2 + \frac{\gamma}{\gamma - 1} \tilde{p} \right)\right] = - \frac{\tilde{\rho} \tilde{u}}{\tilde{r}^2} + \frac{1}{2} \tilde{q}.
\label{eq:dimlessene}
\end{equation}
These can be combined into an entropy equation:
%\begin{equation}
%\frac{d}{d \tilde{r}} \ln \tilde{p} - \gamma \frac{d}{d\tilde{r}} \ln \tilde{\rho} = \frac{\tilde{q} \left(\left(\gamma-1\right) \tilde{\rho} \left(\tilde{u}^2+1 \right) -2 \gamma \tilde{p} \right)}{2 \tilde{p} \tilde{\rho} \tilde{u}} \label{eq:dimles_cons_entr_1}
%\end{equation}
%or in terms of the speed of sound
\begin{equation}
2 \frac{d}{d \tilde{r}} \ln \tilde{c}_s - \left(\gamma -1\right) \frac{d}{d\tilde{r}} \ln \tilde{\rho} = \frac{\gamma \tilde{q} \left(\left(\gamma-1\right) \left(\tilde{u}^2+1 \right) -2 \tilde{c}_s^2 \right)}{2 \tilde{c}_s^2 \tilde{\rho} \tilde{u}} \, .\label{eq:dimles_cons_entr_2}
\end{equation}

\subsection{\label{sec:integral_eqn} Integral Form of the Equations}
Conservation of mass (eq.~\ref{eq:dimlessmass}) and energy (eq.~\ref{eq:dimlessene}) can be integrated to obtain analytic, closed form relations:
\begin{equation}
\tilde{r}^2 \tilde{\rho} \tilde{u} = \frac{\tilde{r}^{3-\eta} - \tilde{r}_{st}^{3-\eta}}{3-\eta}, \label{eq:int_mas} \, 
\end{equation}
\begin{equation}
\frac{1}{2} \tilde{u}^2 - \frac{1}{2} + \frac{\tilde{c}_s^2}{\gamma - 1} + \frac{\tilde{r}^{3-\eta} + \left(2-\eta \right) \tilde{r}_{st}^{3-\eta}- \left( 3-\eta\right) \tilde{r} \tilde{r}^{2-\eta}_{st}}{\left(2 - \eta \right) \tilde{r} \left(\tilde{r}^{3-\eta} - \tilde{r}_{st}^{3-\eta} \right)} = 0 \label{eq:int_enr} \, .
\end{equation}
By substituting these two relations into the third equation, we can reduce the problem to just one first order ordinary differential equation for the Mach number $m = u/c_s$ (Section \ref{sec:gen_sol}).  

Taking the limit $\tilde{u} \rightarrow 0$ and $\tilde{r}\rightarrow \tilde{r}_{st}$ in equation \ref{eq:int_enr} yields the speed of sound at the stagnation point \citep{generozov_et_al_2015}
\begin{equation}
\tilde{c}_s \left( \tilde{r}_{st} \right) = \sqrt{\frac{\gamma - 1}{2}} \, , 
\end{equation}
a result which follows intuitively because the neighbourhood of the stagnation point is close to being in hydrostatic equilibrium.  Taking the limit $\tilde{u} \rightarrow 0$ and $\tilde{r}\rightarrow \tilde{r}_{st}$ in equation \ref{eq:int_mas} also results in an expression for the gas density at the stagnation point
\begin{equation}
\tilde{\rho} \left( \tilde{r}_{st}\right) = \frac{\tilde{r}_{st}^{-\eta}}{\tilde{u}'\left(\tilde{r}_{st} \right)} \, .
\end{equation}
\subsection{Degenerate Cases}

The hydrodynamical equations are greatly simplified when one of the dimensional parameters ($D$, $v_w$ or $M$) vanishes. In this section we discuss each of the three cases. Note that, because one of the dimensional parameters vanishes, we cannot use the dimensionless equations derived in the previous section.

\subsubsection{Bondi Accretion / Parker Wind ($D = 0$)}

Setting the source term $D = 0$ reduces the problem to that of Bondi accretion \citep{bondi_1952} or a Parker wind \citep{parker_1958}.  Absent the injection of mass or energy, entropy is conserved, 
\begin{equation}
S = P/\rho^{\gamma} = \text{constant}
\end{equation} 
and all hydrodynamic equations can be integrated.  Conservation of mass becomes
\begin{equation}
\dot{M} = 4 \pi \rho v r^2 = \text{constant}, 
\end{equation}
while conservation of momentum results in a modified Bernoulli equation
\begin{equation}
B = \frac{1}{2} u^2 + \frac{\gamma}{\gamma-1} \frac{p}{\rho} - \frac{GM}{r} = \text{constant} \, .
\end{equation}
The last equation can also be interpreted as conservation of enthalpy.  These equations constitute an implicit solution for the hydrodynamic profiles.

\subsubsection{Solutions without Gravity ($M = 0$)} \label{sec:sol_wo_gravity}

In this case the gravity of the central black hole is neglected, $M = 0$. The Mach number of the flow varies with radius according to 
\begin{equation}
\frac{r}{r_0}=m^{-\frac{1}{\eta -1}} \left((\gamma -1) m^2+2\right)^{\frac{-\gamma -1}{2 (\gamma  (\eta-5)+\eta -1)}} \times
\end{equation}
\begin{equation*}
\left(1-\eta -\gamma  (\eta -3) m^2\right)^{\frac{\gamma  (\eta -3)+\eta-1}{(\eta -1) (\gamma  (\eta -5)+\eta -1)}}
\end{equation*}
%\begin{equation}
%\frac{r}{r_0} = \left(\gamma-1 +\frac{2}{m^2} \right)^{\frac{2 (\gamma +1)}{\gamma  (3 \eta -7)-\eta +1}} \times
%\end{equation}
%\begin{equation*}
%\left(\gamma  (\eta -3)-\frac{\eta -1}{m^2}  \right)^{\frac{4 (\gamma 
%   (3-\eta)-\eta +1)}{(\eta -1) (\gamma  (3 \eta -7)-\eta +1)}} \, 
%\end{equation*}
where $r_0$ is a constant.  When $1< \eta < 3$, solutions exists for which the Mach number takes on a constant value given by
\begin{equation}
m = \sqrt{\frac{\eta-1}{\gamma \left( 3-\eta\right)}} \, .
\end{equation}
The flow is therefore supersonic for sufficiently steep stellar profiles, 
\begin{equation} \eta > \frac{3 \gamma+1}{\gamma+1} \underset{\gamma = 5/3}= \frac{9}{4}.
\end{equation}
For these $m = constant$ solutions, the velocity and sound speed individually take on constant values:
\begin{equation}
v =  v_w \sqrt{\frac{ \left(\eta-1 \right) \left(\gamma-1 \right)}{2 \left( \gamma-1\right) + \left(3 - \eta \right) \left(\gamma+1 \right)}}, \,,
\label{eq:constv}
\end{equation}
\begin{equation}
c_s =  v_w \sqrt{\frac{\gamma \left(3-\eta \right) \left(\gamma-1 \right)}{2 \left( \gamma-1\right) + \left(3 - \eta \right) \left(\gamma+1 \right)}} \,,
\end{equation}
while the density profile obeys
\begin{equation}
\rho = \frac{\sqrt{2 \left( \gamma-1\right) + \left( 3-\eta\right) \left( \gamma+1\right)}}{\left(3-\eta \right) \sqrt{\eta-1} \sqrt{\gamma-1}} \frac{D}{v_w} r^{1-\eta} \, .
\end{equation}

When $\eta>3$ or $\eta<1$, mass injection must be truncated on small or large radial scales, respectively, to prevent the density profile from diverging.  One way to understand the requirement $\eta>1$ is as follows. From conservation of mass (eq.~\ref{eq:int_mas}), the mass flux scales obeys $\rho v \propto r^{1-\eta}$. When $\eta>1$, the velocity is constant with radius (eq.~\ref{eq:constv}) and so the density decreases towards large radii.

By contrast, for $\eta<1$ the density is constant and so the velocity must increase with radius.  One can understand why the density must be constant in this case by the following argument.  In the limit $r\rightarrow 0$ the flow is hydrostatic, so the pressure is constant, $v \ll c_{s}$ so the speed of sound is also constant $c_{s} \approx v_w$ and therefore the density must be constant and $v\propto r^{\eta-1}$.  In cases for which mass injection is truncated above a certain radius $r_t$ ($q \propto \theta \left(r_t - r\right)$), \citet{chevalier_clegg_1985} show that the sonic point occurs at $r_t$, such that the constant density $\rho_t$ inside $r_t$ can be estimated as
\begin{equation}
\rho_t \approx \frac{D}{v_w} r_t^{1-\eta} \, .
\end{equation}
This makes clear that the density will diverge if $\eta<1$ and the mass injection is not truncated, i.e. $r_t \rightarrow \infty$.

\subsubsection{Slow Mass Injection ($v_{w} = 0$)}

Here we consider the limit that the gas mass is injected with zero velocity $v_{\rm w} = 0$, corresponding to a negligible heating rate. The Mach number profile is given by
\begin{equation}
\frac{r}{r_0} = m^{\frac{2-2 \gamma }{2 \gamma  \eta -3 \gamma -2 \eta +1}} \left((\gamma -1) m^2+2\right)^{\frac{\gamma +1}{-2 \gamma  \eta +9 \gamma -2 \eta
   +1}} \times
\end{equation}
\begin{equation*}
\left(-2 \eta +2 \gamma ^2 (\eta -3) m^2+\gamma  \left(2 \eta -2 (\eta -3) m^2-3\right)+1\right)^{\nu},
\end{equation*}
where $r_0$ is again a constant and
\begin{equation}
\nu \equiv {\frac{4 \gamma ^2 (\eta -3)+6 \gamma -4
   \eta +2}{\gamma ^2 \left(4 \eta ^2-24 \eta +27\right)+6 \gamma  (2 \eta -1)-(1-2 \eta )^2}} \, .
\end{equation}

When $\eta > 3$, solutions with a radially-constant Mach number again exist with
\begin{equation}
m = -\sqrt{\frac{3 \left(\gamma -1\right)+2 \left(2-\eta \right)}{\gamma \left(\gamma-1 \right) \left(3-\eta \right)}} \, .
\end{equation}
The velocity and sound speed individually scale with the free-fall velocity:
\begin{equation}
v = -\frac{\sqrt{-6 \gamma +4 \eta -2}}{\sqrt{(\eta -2) (\gamma  (2 \eta -9)+2 \eta -1)}} \sqrt{\frac{GM}{r}} \, ,
\end{equation}
\begin{equation}
c_s = \sqrt{\frac{2 (\gamma -1) \gamma  (\eta -3)}{(\eta -2) (\gamma  (2 \eta -9)+2 \eta -1)}} \sqrt{\frac{GM}{r}} \, ,
\end{equation}
The gas density profile is given by
\begin{equation}
\rho = \frac{\sqrt{(\eta -2) (\gamma  (2 \eta -9)+2 \eta -1)}}{(3-\eta) \sqrt{-6 \gamma +4 \eta -2}} \frac{D}{\sqrt{GM}} r^{\frac{3}{2}-\eta} \, .
\end{equation}

\subsection{\label{sec:sec_analytic} Analytic Solution}
For the special values $\gamma = 5/3$ and $\eta = 5/2$, we have discovered a completely analytic solution for the hydrodynamic profiles, given by
\begin{equation}
\tilde{\rho} = \frac{4 \sqrt{6}}{3 \tilde{r}^{3/2}} \, ,
\end{equation}
\begin{equation}
\tilde{u} = \frac{\sqrt{6}}{4} - \frac{1}{\sqrt{\tilde{r}}} \, ,
\end{equation}
\begin{equation}
\tilde{c}_s^2 = \frac{5}{24} + \frac{1}{3 \tilde{r}} \, ,
\end{equation}
with the stagnation point located at
\begin{equation}
\tilde{r}_{st} = \frac{8}{3} \, .
\end{equation}
Though this may appear to be a highly-specialised case, an adiabatic index $\gamma = 5/3$ well-approximates that of a non-relativistic plasma when radiative cooling is negligible, while the density of the young stars within the central parsec of the Galactic Centre obey $\eta \approx 2$ \citep{genzel_et_al_2003}.

\subsection{\label{sec:asym_be} Asymptotic Behaviour}

The asymptotic behaviour of the flow depends most sensitively on the value $\eta$. When $\eta<3$, the accretion rate approaching $r\rightarrow 0$ is finite, while the outward mass flux diverges towards $r\rightarrow \infty$ diverges. Well inside the stagnation radius, the solution must therefore converge to the classical Bondi solution, for which the accretion rate is simply $\dot{M} = \int_0^{r_{\rm st}} 4 \pi r^2 \cdot D r^{-\eta} dr = \frac{4 \pi D}{3 - \eta} r_{\rm st}^{3-\eta}$. 

If, on the other hand, $\eta>3$, then the mass per unit time flowing into $r\rightarrow 0$ becomes infinite, while mass flowing to $r\rightarrow \infty$ is finite. Hence, at radii well outside the stagnation radius the solution must converge to that of a Parker wind. The mass loss rate for this case is $\dot{M} = \int_{r_{\rm st}}^{\infty} 4 \pi r^2 D r^{-\eta} dr = \frac{4 \pi D}{\eta - 3} r_{\rm st}^{3-\eta}$. 

The flow properties also depend on the adiabatic index $\gamma$.  Adopting a value $\gamma < 5/3$ is, to some extent, equivalent to introducing cooling to the system.  This is because energy can be transferred to atomic or molecular degrees of freedom that do not increase the pressure.  The Bondi accretion solution is well known to exhibit qualitatively different behaviour depending on whether the value of $\gamma$ is above or below 5/3.

\subsubsection{Limit 1: $r \ll R_b$, $\eta<3$ and $\gamma<5/3$} \label{sec:limit1}
In this limit the velocity is simply that of free-fall,
\begin{equation}
\tilde{u} \approx -\sqrt{\frac{2}{\tilde{r}}} \, .
\end{equation}
From equation \ref{eq:int_mas}, the density profile is then given by
\begin{equation}
\tilde{\rho} \approx \frac{\tilde{r}_{st}^{3-\eta}}{\sqrt{2} \left(3-\eta \right) \tilde{r}^{3/2}}.
\end{equation}
Collision between the mass injected by stars and the infalling flow generates entropy.  If the entropy gained by this process is negligible with respect to the initial value of the inflow on large scales, then the speed of sound will grow with decreasing radius according to
\begin{equation}
\tilde{c_{s}}^2 \propto \tilde{\rho}^{\left(\gamma-1\right)} \propto \tilde{r}^{-3 \left(\gamma-1\right)/2} \, .
\end{equation}
If, on the other hand, the added entropy exceeds its initial value, then, according to equation \ref{eq:dimles_cons_entr_2} we expect
\begin{equation}
\tilde{c_{s}}^2 \propto \tilde{q} \tilde{u} \tilde{r} / \tilde{\rho} \propto \tilde{r}^{2-\eta} 
\label{eq:option2}
\end{equation}
These two cases are distinguished by comparing the exponents.  For a sufficiently steep stellar profile,
\begin{equation}
\eta > \frac{3 \gamma+1}{2} \underset{\gamma = 5/3} = 3,
\end{equation}
the growth in entropy will diverge to small radii and sound speed will grow according to equation (\ref{eq:option2}).

\subsubsection{Limit 2: $r \ll R_b$, $\eta<3$ and $\gamma=5/3$} \label{sec:limit2}

In this case the radial profiles of the velocity and sound speed are given, respectively, by
\begin{equation}
\tilde{u} \approx -\frac{\beta}{\sqrt{\tilde{r}}};  \,\,\,\,\tilde{c}_s \approx \frac{\alpha}{\sqrt{\tilde{r}}} \, ,
\end{equation}
while from equation \ref{eq:int_mas} the density profile is given by
\begin{equation}
\tilde{\rho} \approx \frac{\tilde{r}_{st}^{3-\eta}}{\left(3-\eta \right) \tilde{r}^{3/2} \beta},
\end{equation}
where $\alpha$ and $\beta$ are constants.  Constraints on the values of $\alpha$ and $\beta$ are obtained from equation \ref{eq:int_enr}
\begin{equation}
\frac{\tilde{r} \tilde{c}_s^2}{\gamma - 1} + \frac{\tilde{r} \tilde{u}^2}{2} \approx \frac{3}{2}\alpha^2 + \frac{\beta^2}{2}\approx 1 \, . \label{eq:alpha_beta_1}
\end{equation}
and equation \ref{eq:dimles_cons_entr_2},
\begin{equation}
3 \alpha^2 - \beta^2 = 0 \, . \label{eq:alpha_beta_2}
\end{equation}
Solving equations \ref{eq:alpha_beta_1} and \ref{eq:alpha_beta_2} yields
\begin{equation}
\alpha = \frac{1}{\sqrt{3}}; \,\,\,\,\beta = 1,
\end{equation}
in agreement with our analytic solution (Section \ref{sec:sec_analytic}).

\subsubsection{Limit 3: $r \ll R_b$, $\eta<3$ and $\gamma > 5/3$}
In this case thermal energy dominates the energy equation, such that
\begin{equation}
\tilde{c}_s \approx \sqrt{\frac{\gamma - 1}{\tilde{r}}}.
\end{equation}
and the density profile is given by equation \ref{eq:dimles_cons_entr_2}
\begin{equation}
\rho = \tilde{\rho} \left( \tilde{r}_{st} \right) \left( \frac{\tilde{r}_{st}}{\tilde{r}}\right)^{\frac{1}{\gamma - 1}}.
\end{equation}
The velocity profile then follows from equation \ref{eq:int_mas}
\begin{equation}
\tilde{u} \approx - \frac{\tilde{r}_{st}^{3-\eta}}{2^{\frac{1}{\gamma-1}}\left( 3 -\eta\right) \tilde{\rho} \left( \tilde{r}_{st}\right)} r^{\frac{3-2\gamma}{\gamma-1}}
\end{equation}

\subsubsection{Limit 4: $r \gg R_b$ and $\eta<3$} \label{sec:rinf_etalt3}
In this limit, gravity is negligible and the flow converges to the solution given in section \ref{sec:sol_wo_gravity}. The sound speed and the fluid velocity are radially constant
\begin{equation}
c_{s} \approx v_{w} \sqrt{\frac{\gamma \left(\eta - 3\right) \left(\gamma - 1\right)}{\eta \gamma + \eta - 5 \gamma - 1}};
\end{equation}
\begin{equation}
u \approx v_{w} \sqrt{- \frac{\left(\eta - 1\right) \left(\gamma - 1\right)}{\eta \gamma + \eta - 5 \gamma - 1}}.
\end{equation}
At large distances the Mach number approaches a constant asymptotic value,
\begin{equation}
m \approx \sqrt{\frac{\eta-1}{\gamma \left(3-\eta \right)}}. \label{eq:asym_mach_number}
\end{equation}
The density follows a power-law radial profile,
\begin{equation}
\rho \approx \frac{D \sqrt{5 \gamma + 1 - \eta \left(\gamma - 1\right) - 2 \eta }}{v_{w} \left(3-\eta \right) \sqrt{\eta - 1} \sqrt{\gamma - 1}} r^{-\eta+1}. \label{eq:rho_rinf_etalt3}
\end{equation}
The flow becomes supersonic at large distances for 
\begin{equation}
\eta > \frac{3 \gamma + 1}{\gamma + 1} \underset{\gamma =5/3}= \frac{9}{4}
\end{equation}
The Mach number is only real if $1 < \eta < 3$. If $\eta>3$, then the mass generated inside any finite radius is infinite and so the density must diverge at large radii.  When $\eta<1$ no steady-state solution exists, a case we do not treat in this work. 

\subsubsection{Limit 5: $r \ll R_b$ and $\eta>3$} \label{sec:limit5}
In this case, stellar mass injection is dominated by small radii and so the adiabatic relations no longer apply. The velocity and speed of sound both vary as the free-fall velocity $\propto r^{-1/2}$, with prefactors determined from equations \ref{eq:dimles_cons_entr_2} and \ref{eq:int_enr}
\begin{equation}
\tilde{u} = -\frac{\sqrt{-6 \gamma +4 \eta -2}}{\sqrt{\eta -2} \sqrt{\gamma  (2 \eta -9)+2 \eta -1}} \frac{1}{\sqrt{\tilde{r}}}
\end{equation}
\begin{equation}
\tilde{c}_s = \sqrt{2} \sqrt{\frac{(\gamma -1) \gamma  (\eta -3)}{(\eta -2) (\gamma  (2 \eta -9)+2 \eta -1)}} \frac{1}{\sqrt{\tilde{r}}}.
\end{equation}
The density profile follows from equation \ref{eq:int_mas}
\begin{equation}
\tilde{\rho} = \frac{\sqrt{\eta -2} \sqrt{\gamma  (2 \eta -9)+2 \eta -1} }{(\eta -3) \sqrt{-6 \gamma +4 \eta -2}} \tilde{r} ^{\frac{3}{2}-\eta } \, .
\end{equation}
The mach number is given by
\begin{equation}
m \approx \sqrt{\frac{2}{\gamma-1}} > 1,
\end{equation}
such that this asymptotic is always supersonic.

\subsubsection{Limit 6: $r\gg R_b$ and $\eta>3$} \label{sec:limit6}
In this case, most of the mass in the wind is generated close to the stagnation point and the solution coincides with a Bondi wind.  The terminal speed of sound vanishes and, using  equation \ref{eq:int_enr},  the  velocity profile is given by
\begin{equation}
\tilde{u}^2 \approx 1 - \frac{2 \left(\eta - 3 \right)}{\left( \eta - 2\right) \tilde{r}_{st}}
\end{equation}
The density profile is then given by equation \ref{eq:int_mas}
\begin{equation}
\tilde{\rho} \approx \frac{\tilde{r}_{st}^{3-\eta}}{\left(\eta - 3 \right) \tilde{r}^2 \tilde{u}}
\end{equation}
Finally, the sound speed profile is given by
\begin{equation}
 \tilde{c}_s^2 \approx \frac{\gamma  (\eta -3) \tilde{r}_{st}^{\eta -4} (\eta  \tilde{r}_{st}-2 \eta -2 \tilde{r}_{st}+6)}{(\eta -2) (\eta -1) \tilde{r}^{\eta - 3}}  +\frac{\gamma }{3 \tilde{r} }
\end{equation}

\subsubsection{Summary of Flow Properties}
\begin{figure}
\includegraphics[width=0.9\columnwidth]{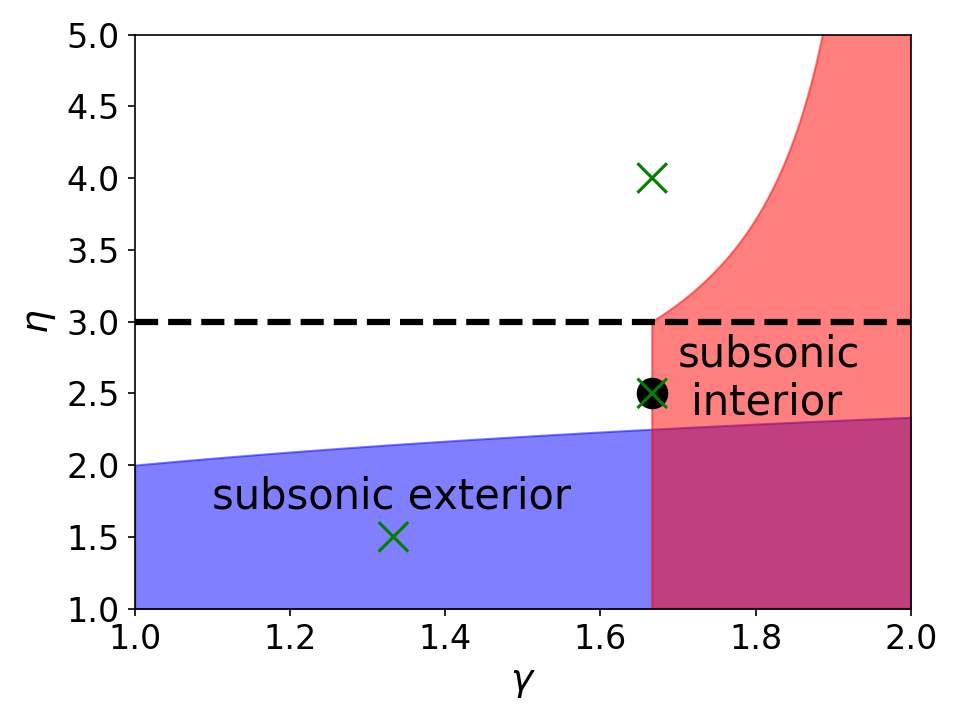}
%\captionsetup{justification=raggedright,
%singlelinecheck=false
%}
\caption{
Phase diagram for the generalised Bondi problem in the space of adiabatic index $\gamma$ and radial power-law index $\eta$ of the mass-injection source ($q(r) \propto r^{-\eta}$; eq.~\ref{eq:eta}).  Shaded regions denote the space of parameters for which the flow asymptotes to a subsonic velocity at radii $r \ll R_b$ (red) and $r \gg R_b$ (blue), where $R_{\rm b}$ is the Bondi radius.  A dashed line denotes the critical value of $\eta = 3$, separating regions for which most of the gaseous mass is generated on large radial scales $r \rightarrow \infty$ ($\eta<3$) versus small radial scales $r \rightarrow 0$ ($\eta > 3$).  A black dot denotes the parameters corresponding to the fully analytic solution (Section \ref{sec:sec_analytic}).  Crosses denote the parameter values used in our numerical simulations presented in Section \ref{sec:numerical_sim}.
\label{fig:phase_diagram}
}
\end{figure}

The results of this section are summarised by Figure \ref{fig:phase_diagram}, which shows a ``phase diagram'' in the space of adiabatic index $\gamma$ and mass source term power-law index $\eta$.  Throughout much of the parameter space, the flow on both small and large radial scales becomes supersonic, resulting in all three critical points being achieved across the domain.  However, for $\gamma > 5/3$ there are regions where the inflow remains subsonic at small radii ($r \ll R_b$), and others where the outflow remains subsonic at large radii ($r \gg R_{\rm b}$).  Even for $\gamma < 5/3$, the outflow will remain subsonic for $\eta \lesssim 2$.

\subsection{Truncation of the Mass Source Term} \label{sec:trunction}

In the inner parsec of the galactic centre, gas heating is dominated by powerful stellar winds from a cluster of massive stars.  The density of the massive stars is decreases sharply at radial distances larger than 0.4 pc from SgrA* \citep{genzel_et_al_2010}.  In the previous sections, we showed that for some values of $\eta$, no steady state solution exists if the mass injection occurs uniformly with radius.  However, if the source of mass injection is truncated above a certain radius, as approximately realised in our own Galactic Centre, then a steady-state solution can still be achieved.  

Truncated solutions have previously been considered analytically for the case $\eta=0$ neglecting gravity \citep{chevalier_clegg_1985}, and numerically for a few values of $\eta$ with gravity included \citep{quataert_2004}. In this section we aim to provide a more complete description, as well as an intuitive explanation to the results in those studies.  For pedagogic purposes, we begin with an abstract exercise in fluid dynamics, and gradually introduce details, slowly working our way toward a model for a galactic centre. 

First consider a case where mass injection is concentrated in a hollow spherical shell of radius $R_o$, i.e. $q \propto \delta \left(r - R_o\right)$. Suppose the net mass injection rate is $\dot{M}$, and that the energy per unit mass is $v_w^2$. Inside the shell $r<R_o$ the velocity is zero, the density is constant $\dot{M}/v_w R_o^2$ and also the pressure $\dot{M} v_w /R_o^2$. Far away from the shell $r\gg R_o$ the velocity is constant $v_w$ and density decreases $\propto r^{-2}$. Across the shell itself there is a discontinuous increase in the velocity from zero to a fraction of $v_w$, followed by an additional increase by a factor of order unity from $r=R_o$ to $r\rightarrow \infty$.

Next, let us consider a case of a uniform mass injection per unit volume between two radii $R_o \gg R_i$. The net mass injection rate is still $\dot{M}$. The interior $r<R_i$ and exterior $r>R_o$ solutions are the same as in the previous case. Inside the mass generating region $R_{i} \ll r \ll R_o $, the gas is almost at hydrostatic equilibrium, such that the speed of sound is $v_w$, the pressure is constant $\dot{M} v_w/R_o^2$, and so the density is constant $\dot{M}/v_w R_o^2$. Conservation of mass mandates that the velocity must increase linearly with radius $v \approx v_w \left(r-R_i \right)/R_o$. 

Next, we consider a case where between $R_o$ and $R_i$ the mass injection rate varies with radius  $\propto r^{-\eta}$. When $\eta<1$, the intermediate region $R_{i} \ll r \ll R_o$ is still nearly hydrostatic, and therefore the velocity scales as $v \propto r^{1-\eta}$. In the case when $\eta>1$ the velocity is constant, and the density declines as $\rho \propto r^{1-\eta}$. In both cases $\rho v \propto r^{1-\eta}$.

Finally, we introduce a gravitating mass $M$ such that the Bondi radius is much smaller than any other length scale in the problem $R_i \gg G M/v_w^2$. Deep inside the Bondi radius $r\ll G M/v_w^2$ the velocity scales with radius as $r^{-1/2}$ and the density as $r^{-3/2}$. The stagnation point is where the velocity in the solution without gravity is comparable to the escape velocity from $M$ at the same distance. The stagnation radius therefore slightly exceeds $R_i$.

\subsection{\label{sec:gen_sol} General Solution to the Non-Truncated Flow Equations}
The governing equations \ref{eq:mass}-\ref{eq:energy} can be reduced to a single ordinary differential equation for the mach number $m$ as a function of the dimensionless radius $\tilde{r} = r/\tilde{r}_{s}$,
\begin{equation}
\label{eq:macheq}
m' = -\frac{m}{4f} \tilde{r}^{-2} \frac{(\gamma -1) m^2+2}{m^2-1} \times 
\end{equation}
\begin{equation*}
\left[\tilde{r}^2 \left(\gamma  m^2+1\right) f'- 4 \tilde{r} f-\gamma  m^2-
\frac{2 \gamma}{\gamma -1}
\right]-
\end{equation*}
\begin{equation*}
m \tilde{r}^{-\eta} \left(\frac{\gamma-1}{2} m^2+1\right) \frac{\gamma  m^2+\tilde{r}^2 }{\left(m^2-1\right) h},
\end{equation*}
where
\begin{equation}
f\equiv \frac{\tilde{r}^{3-\eta }-(3-\eta ) \tilde{r} \tilde{r}_{st}^{2-\eta }+(2-\eta ) \tilde{r}_{st}^{3-\eta }}{(2-\eta ) \tilde{r} \left(\tilde{r}^{3-\eta }-\tilde{r}_{st}^{3-\eta }\right)}-\frac{1}{2}
\end{equation}
\begin{equation}
h\equiv \frac{\tilde{r}^{3-\eta }-\tilde{r}_{st}^{3-\eta }}{3-\eta }
\end{equation}
and a prime indicates a derivative with respect to $\tilde{r}$.

The main challenge to obtaining the solution for an arbitrary choices of parameters is to determine the location of the stagnation point. Once $\tilde{r}_{s}$ is known, equation \ref{eq:macheq} can be numerically integrated to obtain $m(\tilde{r}$).  Then, the spatial profiles of the hydrodynamic variables are obtained from the conserved quantities. 
Equation \ref{eq:macheq} has singularities at the sonic points where $m^2 = 1$ and at the stagnation point $m = 0$. For a given choice of $\tilde{r}_{st}$, the requirement that the mach number pass smoothly through the sonic points dictates their positions. From each sonic point it is possible to numerically integrate the Mach number to the stagnation point. For the wrong choice of $\tilde{r}_{st}$, the curves from each side will not connect smoothly at the stagnation point. Hence, the condition for determining $\tilde{r}_{st}$ is that both numeric integration curves connect smoothly at the stagnation point.
If only an internal sonic point exists, then the integration is instead started from a sufficiently large radius at the Mach number given from equation \ref{eq:asym_mach_number}.

\begin{figure}
    \begin{subfigure}[b]{0.4\textwidth}
        \includegraphics[width=\textwidth]{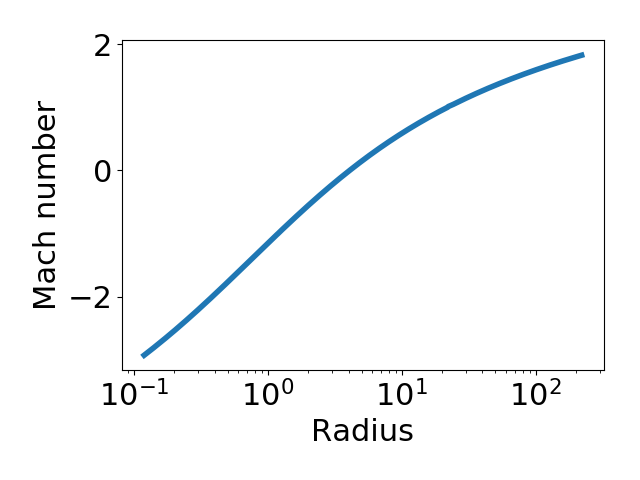}

%        \caption{A gull}
%        \label{fig:gull}
    \end{subfigure}
    ~ %add desired spacing between images, e. g. ~, \quad, \qquad, \hfill etc. 
      %(or a blank line to force the subfigure onto a new line)
    \begin{subfigure}[b]{0.4\textwidth}
        \includegraphics[width=\textwidth]{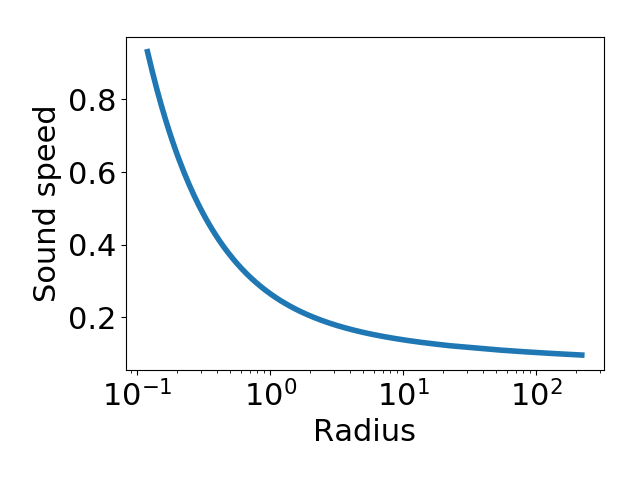}
%        \caption{A tiger}
%        \label{fig:tiger}
    \end{subfigure}
    ~ %add desired spacing between images, e. g. ~, \quad, \qquad, \hfill etc. 
    %(or a blank line to force the subfigure onto a new line)
    \begin{subfigure}[b]{0.4\textwidth}
        \includegraphics[width=\textwidth]{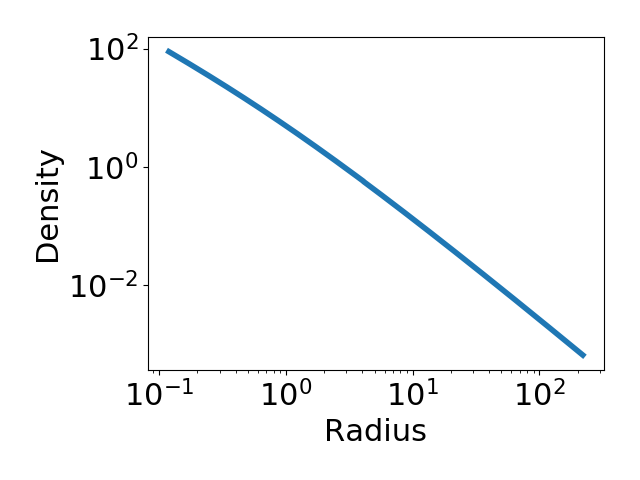}
    \end{subfigure}
    \caption{Steady-state radial profiles for the Mach number (top panel), dimensionless sound speed (middle panel) and dimensionless density (bottom panel) as a function of dimensionless radius $\tilde{r}_{st}$, calculated for $\gamma = 4/3$ and $\eta = 2.9$.  Behaviour at small radii is discussed in \ref{sec:limit1} and at large radii in section \ref{sec:rinf_etalt3}.}
    \label{fig:case_1}
\end{figure}

\begin{figure}
    \begin{subfigure}[b]{0.4\textwidth}
        \includegraphics[width=\textwidth]{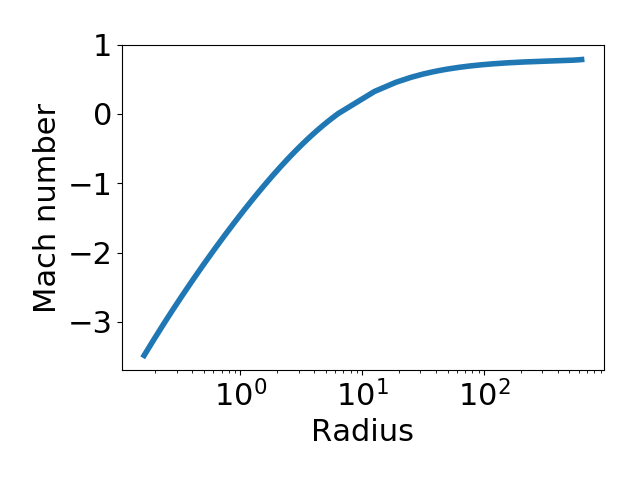}
    \end{subfigure}
    \begin{subfigure}[b]{0.4\textwidth}
        \includegraphics[width=\textwidth]{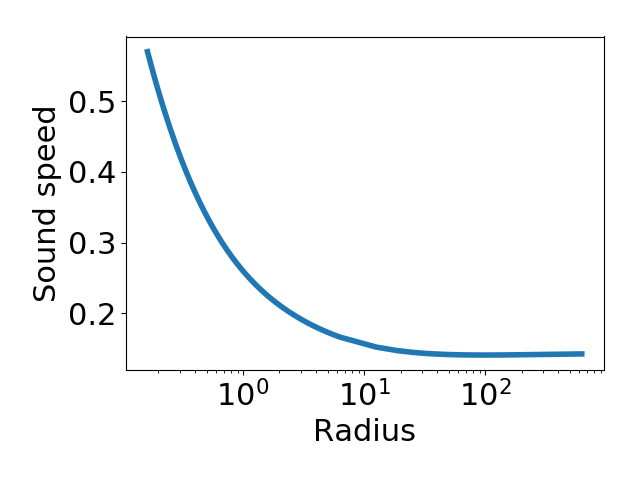}
    \end{subfigure}
    \begin{subfigure}[b]{0.4\textwidth}
        \includegraphics[width=\textwidth]{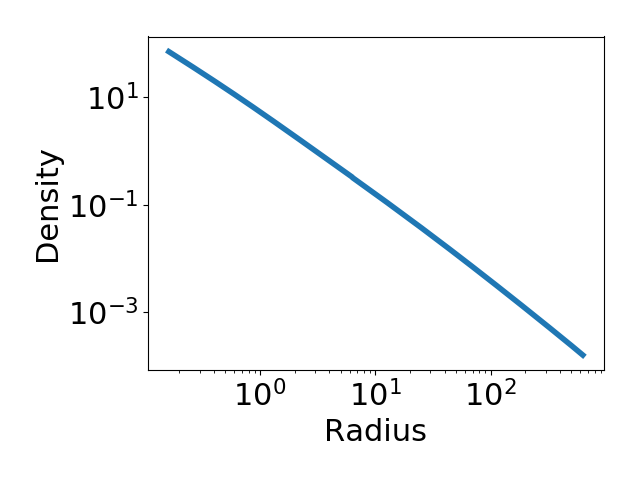}
    \end{subfigure}
    \caption{Same as Figure \ref{fig:case_1}, but calculated for $\gamma = 4/3$, $\eta = 1.9$. 
    }\label{fig:case_2}
\end{figure}

\begin{figure}
    \centering
    \begin{subfigure}[b]{0.4\textwidth}
        \includegraphics[width=\textwidth]{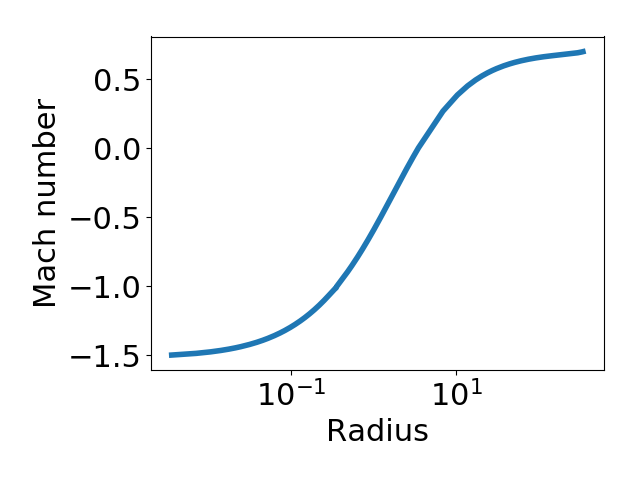}
    \end{subfigure}
    \begin{subfigure}[b]{0.4\textwidth}
        \includegraphics[width=\textwidth]{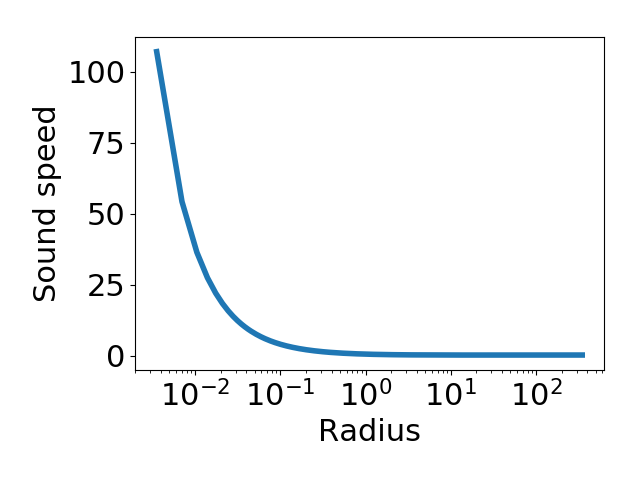}
    \end{subfigure}
    \begin{subfigure}[b]{0.4\textwidth}
        \includegraphics[width=\textwidth]{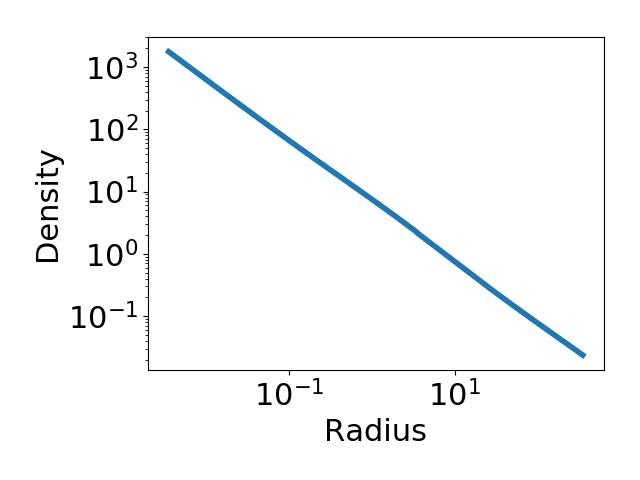}
    \end{subfigure}
    \caption{Same as Figure \ref{fig:case_1}, but calculated for $\gamma = 5/3$, $\eta = 1.9$. Behaviour at small radii is discussed in \ref{sec:limit2}.
}\label{fig:case_3}
\end{figure}

Figures \ref{fig:case_1} and \ref{fig:case_2} the hydrodynamic profiles for two examples in which the flow becomes supersonic and subsonic, respectively, at large distances.  Note also that there is a qualitative difference between cases for which $\gamma = 5/3$ and $\gamma < 5/3$. In the latter case, the flow energy sufficiently close to the central mass becomes purely kinetic and the mach number diverges. However, when $\gamma = 5/3$, as illustrated in Fig.~\ref{fig:case_3}, the ratio between the kinetic and thermal energies is radially constant, as is the Mach number. 

\begin{figure}
\includegraphics[width=0.9\columnwidth]{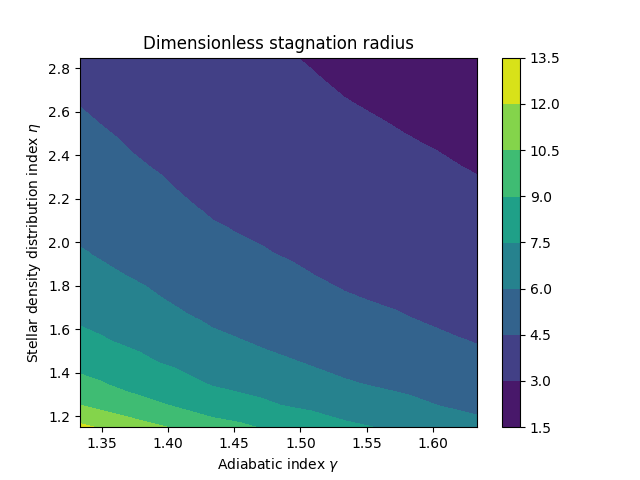}
\caption{
Contours of the dimensionless stagnation radius $\tilde{r}_{st} = r_{\rm st}/R_{\rm b}$, where $R_{\rm b}$ is the standard Bondi radius, as a function of the adiabatic index, $\gamma$, and the radial slope of the stellar density profile, $\eta$.  The value of $\tilde{r}_{st}$ does not vary appreciably over the parameter space of astrophysical interest, except for a divergence in the limit $\eta \rightarrow 1$; this reflects the fact that no steady state, non-truncated solutions exist in the region $\eta<1$ (see Section \ref{sec:rinf_etalt3}).
}
\label{fig:r_st_map}
\end{figure}

Figure \ref{fig:r_st_map} shows a contour map of the dimensionless stagnation radius $\tilde{r}_{st}$ as a function of the adiabatic index, $\gamma$, and the slope of the mass source term, $\eta$. Over the range of parameters relevant to astrophysical problems, the dimensionless stagnation radius is of order unity, i.e. within a factor of $\lesssim 10$ times the usual Bondi radius.  However, the value of $\tilde{r}_{st}$ diverges as $\eta \rightarrow 1$, reflecting the fact that no steady, non truncated solutions exist for $\eta<1$ (see Section \ref{sec:rinf_etalt3}).  For $\gamma=5/3$ and $1.1 < \eta < 3$, \cite{generozov_et_al_2015} show that the value of $\tilde{r}_{st}$ is well-approximated by the following expression,
\begin{equation}
\tilde{r}_{st} \approx \frac{6}{\eta-1/4}
\end{equation}

\section{Numerical Simulation} \label{sec:numerical_sim}
To verify our analytic solutions, we have performed a series of one dimensional, time-dependent simulations. We solve the time dependent equations in terms of primitive variables ($\log(\rho)$, $v$, and $s$) using a sixth order difference scheme for spacial derivatives, and a third order Runge-Kutta scheme for time integration. We add artificial viscosity terms to the velocity and entropy equations for numerical stability \citep{brandenburg_2003, generozov_et_al_2015}. The grid is logarithmic with 100 grid points. We perform power law extrapolations of the flow properties at the boundaries. We ran these simulations until a steady state was reached

\begin{figure}
    \centering
    \begin{subfigure}[b]{0.4\textwidth}
        \includegraphics[width=\textwidth]{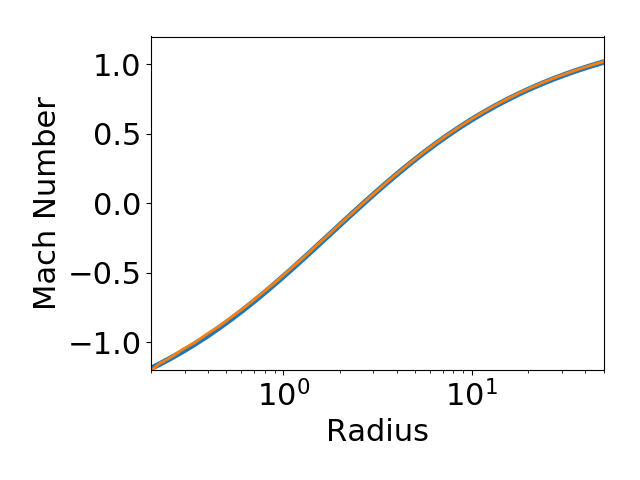}
    \end{subfigure}
    \begin{subfigure}[b]{0.4\textwidth}
        \includegraphics[width=\textwidth]{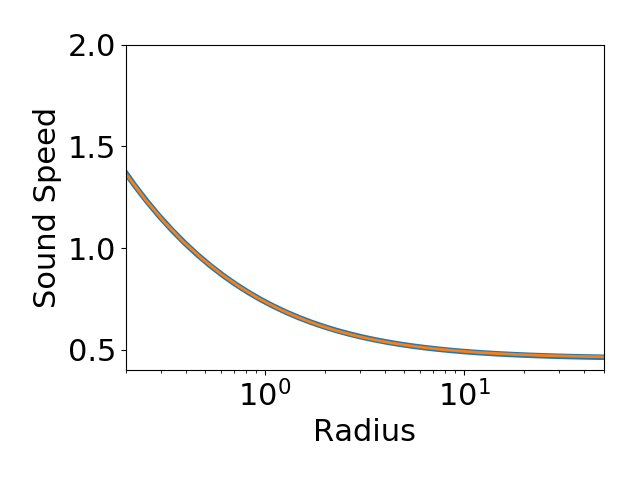}
    \end{subfigure}
    \begin{subfigure}[b]{0.4\textwidth}
        \includegraphics[width=\textwidth]{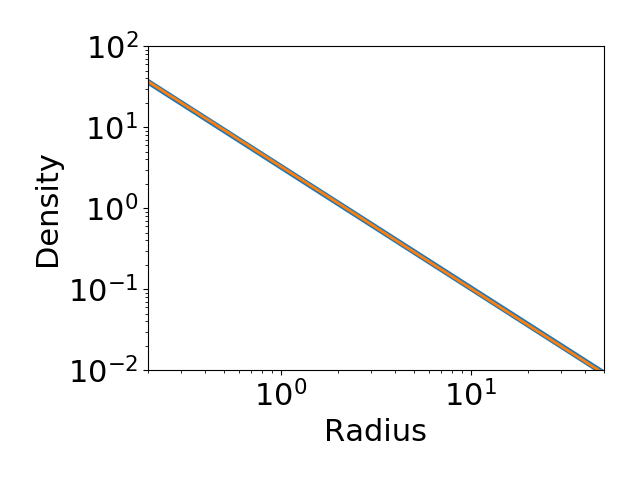}
    \end{subfigure}
    \caption{Snapshot of the time-dependent numerical solutions (orange) for the radial profiles of density, sound speed, and velocity, calculated for $\gamma = 5/3$ and $\eta = 5/2$.  Shown for comparison is the analytic steady-state solution (blue). Behaviour at small radii is discussed in \ref{sec:limit2} and at large radii in section \ref{sec:rinf_etalt3}.
\label{fig:simulation_comparison}
}
\end{figure}

\begin{figure}
    \centering
    \begin{subfigure}[b]{0.4\textwidth}
        \includegraphics[width=\textwidth]{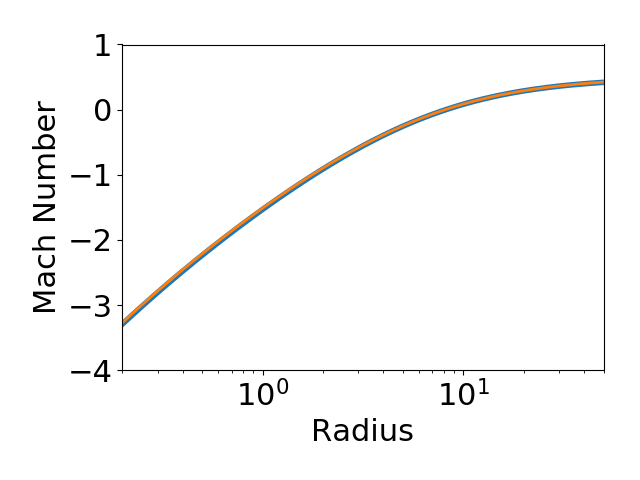}
    \end{subfigure}
    \begin{subfigure}[b]{0.4\textwidth}
        \includegraphics[width=\textwidth]{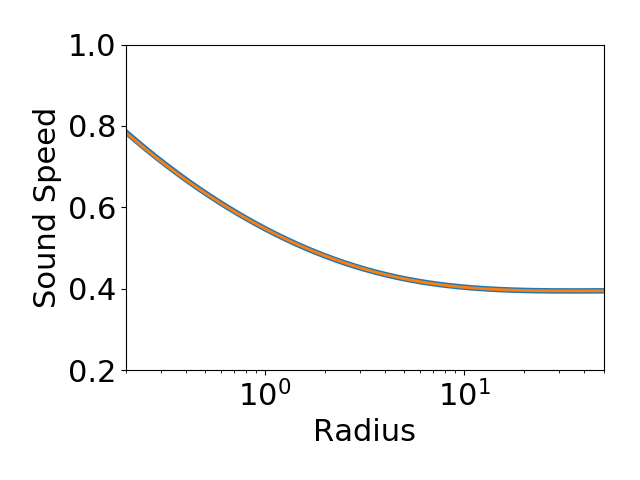}
    \end{subfigure}
    \begin{subfigure}[b]{0.4\textwidth}
        \includegraphics[width=\textwidth]{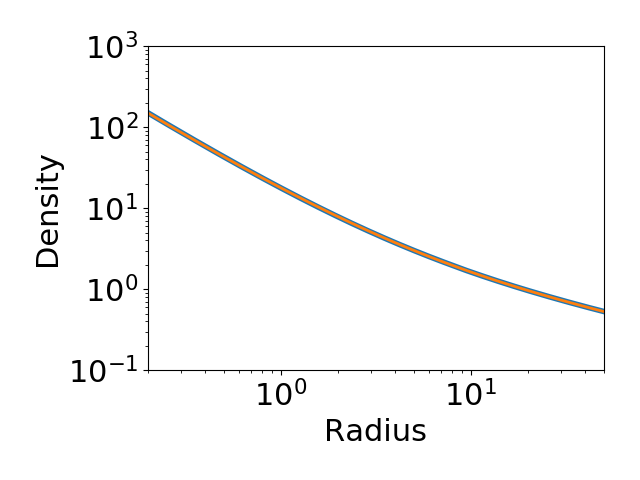}
    \end{subfigure}
    \caption{Same as Figure \ref{fig:simulation_comparison}, but for $\gamma = 4/3$ and $\eta = 3/2$. Behaviour at small radii is discussed in \ref{sec:limit1}.
\label{fig:simulation_comparison_2}
}
\end{figure}

\begin{figure}
    \centering
    \begin{subfigure}[b]{0.4\textwidth}
        \includegraphics[width=\textwidth]{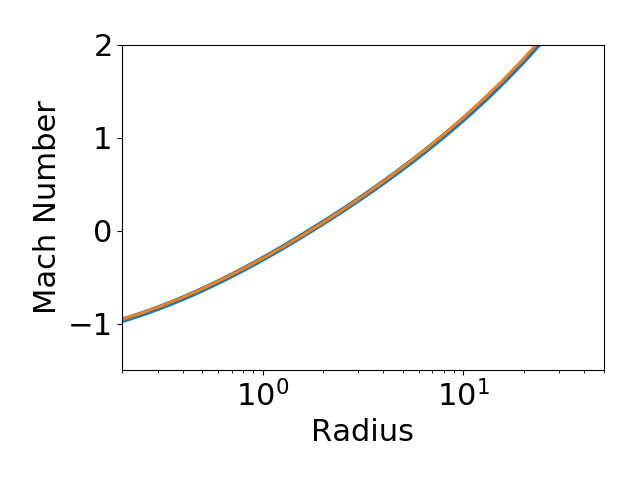}
    \end{subfigure}
    \begin{subfigure}[b]{0.4\textwidth}
        \includegraphics[width=\textwidth]{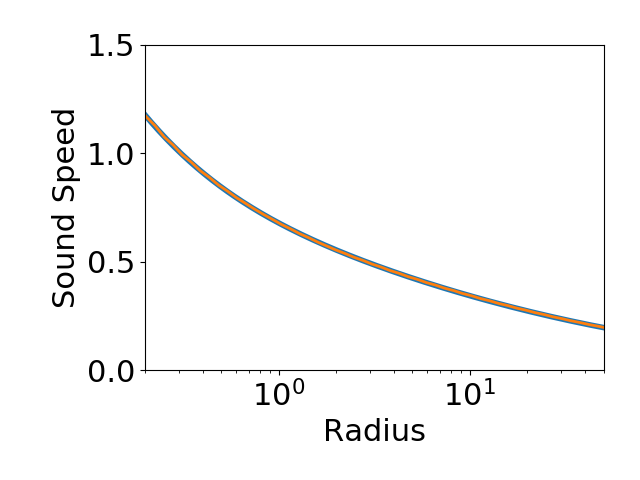}
    \end{subfigure}
    \begin{subfigure}[b]{0.4\textwidth}
        \includegraphics[width=\textwidth]{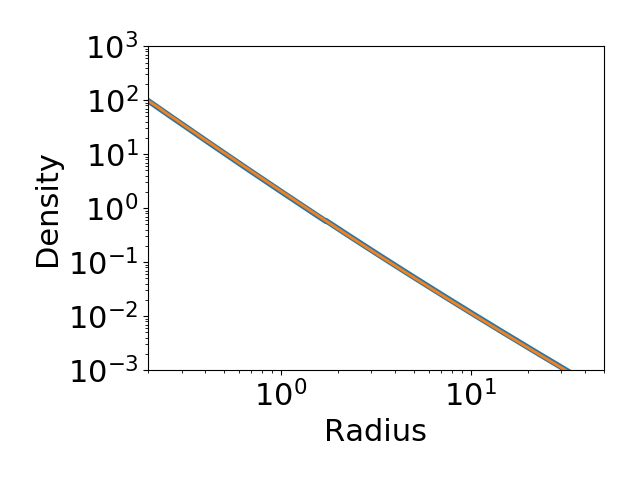}
    \end{subfigure}
    \caption{Same as Figure \ref{fig:simulation_comparison}, but for $\gamma = 5/3$ and $\eta = 4$. Behaviour at small radii is discussed in \ref{sec:limit5} and at large radii in section \ref{sec:limit6}.
\label{fig:simulation_comparison_3}
}
\end{figure}

Mass and energy are injected into each cell, both at the prescribed rate $\propto r^{-\eta}$, for three values of $\eta = 2.5, 1.5,$ and 4. The adiabatic indices used in these cases were $\gamma=$ 5/3, 4/3 and 5/3 respectively.  These parameters were chosen such that the calculations will span different regions of the  relevant space (see Figure \ref{fig:phase_diagram}).  In the first calculation the flow becomes supersonic at both radial extremes ($r\rightarrow0$ or $r\rightarrow \infty$).  In the second, the flow is subsonic at large radii, $r\rightarrow \infty$. In the third calculation, both extremes are subsonic, but $\eta>3$.  The final snapshots of the flow, along with a comparison to the analytic solutions, are presented in Figures \ref{fig:simulation_comparison},\ref{fig:simulation_comparison_2} and \ref{fig:simulation_comparison_3}.

\section{Radiation} \label{sec:radiation}
Our calculations thus far have neglected the effects of radiative cooling, at least explicitly.  In this section we discuss the conditions under which this is valid.  We also calculate the luminosity and spectra of the flow due to free-free emission in this limit.

\subsection{Radiative Cooling} \label{sec:rad_cool}
Radiative cooling contributes a sink term to the entropy equation of the form $\rho^2 \Lambda \left(T\right)$, where $\Lambda(T)$ is the cooling function; this general form accounts for any two-body collisional process, typically including free-free emission and line cooling \citep{sutherland_dopita_1993}.

In order for radiative cooling to appreciably alter the thermal content at a given point in the flow, two conditions must be met. First, the cooling rate should be comparable to the heating rate from stellar winds, i.e.
\begin{equation}
\frac{\rho^2}{m_p^2}\Lambda \gtrsim q v_w^2,
\label{eq:cooling_cond_2}
\end{equation}
where $m_p$ is the proton mass.  Second, the fluid element must have sufficient time to radiate a sizable fraction of its energy before being advected to a different radius.  In other words the cooling timescale $\sim m_p^2 c_s^2/\rho \Lambda$ must be shorter than the advection time $\sim r/u$, i.e.
\begin{equation}
\frac{r}{|u|} > \frac{m_p^2 c_s^2}{\rho \Lambda} \, .
\label{eq:cooling_cond_1}
\end{equation} 

One effect of cooling, when the above conditions are satisfied, is to give rise thermal instabilities \citep{generozov_et_al_2015}, in which some of the gas undergoes runaway cooling and condenses' into dense clumps. At high temperatures, thermal Bremsstrahlung is the dominant cooling mechanism and the $\Lambda \propto T^{1/2}$ dependence of the cooling function gas is thermally stable.  However, at lower temperatures $\lesssim 10^7$ K, line emission comes to dominate, and the gas becomes unstable to runaway cooling (e.g.~\cite{mccourt_et_al_2012}).

On the other hand, even in cases when the thermal energy is strongly depleted by cooling, the general characteristics of the flow do not necessarily change appreciably.  For instance, if the flow is already highly supersonic, then only a small fraction of the energy is in thermal form to begin with, and changes introduced by radiative cooling in supersonic regions are not able to propagate to the subsonic regions. 

\subsection{Free-Free Emission}
%In this section we discuss a few general characteristics of the electromagnetic radiation from the inflow-outflow gas flow.

%\subsubsection{Optical depth}
%Even if the wind profiles extends to infinity, the optical depth does not necessarily diverge. Assuming the opacity is dominated by Compton scattering, the optical depth at great distances from black hole in the case $\eta<3$ scales as $\tau \propto \rho r \propto r^{2-\eta}$ (see equation \ref{eq:rho_rinf_etalt3}). Therefore, the optical depth diverges only if $\eta<2$. In the case $\eta>3$, the mass at large distances is dominated by winds close to the stagnation radius, so the optical depth scales as $\tau \propto \rho r \propto r^{-1}$ so the optical depth does not diverges.

We now calculate the free-free emission of the flow, focusing on the case $\eta = 5/2$ and $\gamma = 5/3$ for which we have obtained a fully analytic solution (section \ref{sec:sec_analytic}).  The contribution to the bolometric luminosity of the flow on radial scales $\sim r$ is given by $L \sim r^3 \Lambda \rho^2 /m_p^2$.  For free-free emission $\Lambda \left( T\right) \propto \sqrt{T}$. The luminosity diverges in the limit $r \ll R_b$ because the density scales as $\rho \propto r^{-3/2}$ and the temperature as $T \propto r^{-1}$ (section \ref{sec:limit2}). In the limit $r \gg R_b$ the density scales as $\rho^{1-\eta}$ and the temperature is constant \ref{sec:rinf_etalt3} so the luminosity diverges at large radii if $\eta<5/2$.  The bolometric luminosity is therefore determined by the radii $r_{\min}$ and $r_{\max}$ at which the stellar distribution, and thus the injection of mass and energy, is truncated.  

\begin{figure}
\includegraphics[width=0.9\columnwidth]{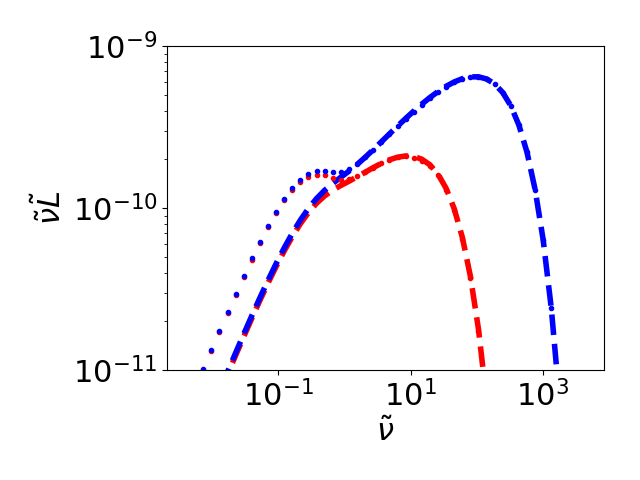}
\caption{
Synthetic spectrum due to free free emission from a flow with $\gamma=5/3$ and $\eta=5/2$. We normalised the frequency such that $\tilde{\nu} = h \nu / m_p v_w^2$ and the luminosity by the accretion power $\tilde{\nu} \tilde{L} = \nu L / \dot{M} c^2$, using values for our galactic centre. Dots represent truncation at $r_{\max} = 1000 \cdot R_b$, and dashes at $r_{\max} = 10 \cdot R_b$. Blue represents truncation at $r_{\min} = R_b/1000$ and red represents truncation at $r_{\min} = R_b/100$. 
\label{fig:synthetic_spectra}
}
\end{figure}

The specific luminosity at frequency $\nu$ is given by
\begin{equation}
L_{\nu} \approx \frac{m_e c^2 r_e^4}{m_p^2} \int_{r_{\min}}^{r_{\max}} \rho^2 \sqrt{\frac{m_e c^2}{k T}} \exp \left(-\frac{h \nu}{k T} \right) r^2 dr. \label{eq:spectral_luminosity}
\end{equation}
At the lowest frequencies, the exponent in the integrand is constant and the spectrum is flat.  By contrast, at the highest frequencies, the exponent decreases so rapidly that only the inner most shell dominates the flow and the spectrum declines exponentially with frequency. At intermediate frequencies, a single radial shell dominates the emission at each frequency and we have $L_{\nu} \propto \nu^{-1/2}$.  These three regimes can be seen in Figure \ref{fig:synthetic_spectra}, which shows the result of integrating \ref{eq:spectral_luminosity} using our analytic solution $\rho(r)$ and $T(r)$ (section \ref{sec:sec_analytic}).

In practice, our result can be used to translate an observed $\nu_0 L_{\nu_0}$ at a given X-ray energy
\begin{equation}
h\nu_0 \equiv m_p v_{\rm w}^{2} \simeq \,10\,{\rm keV}\,\left(\frac{v_{\rm w}}{10^{3}\,{\rm km\,s^{-1}}}\right)^{2}
\end{equation}
into an estimate of the accretion rate onto the SMBH
\begin{equation}
\dot{M} = 7.5 \cdot 10^{-4} M_{\odot}\,{\rm yr^{-1}} \times
\label{eq:Mdotcalib}
\end{equation}
\begin{equation*}
\left(\frac{\nu_0 L_{\nu_0}}{10^{38}\,{\rm erg\,s^{-1}}}\right)^{1/2} \left(\frac{M_h}{4\cdot 10^6 M_{\odot}}\right)^{1/2} \left(\frac{v_w}{500 \, \rm km/s}\right)^{-1}
\end{equation*}

%{\bf BDM: fill this in with formula motivated by the analytic result.  Reference should be made only to observables, assuming they measure the spectrum given in Figure 9.}.

\section{Astrophysical Applications} \label{sec:astro_app}
\subsection{Our Galactic Centre}

The hot gas in the central parsec of our Galactic Centre is continuously supplied by winds from young and massive stars. \citet{quataert_2004} numerically integrated the steady-state hydrodynamic equations using a truncated mass injection term to obtain the density and temperature profiles.  He compared his theoretical results to observational data obtained by {\it Chandra} \citep{baganoff_et_al_2003} and arrived at two contradicting results.  The {\it Chandra} measurements of the density are best fit by a model with $\eta = 0$, while the surface brightness is best fit by $\eta=3$. In this section we wish to analyze the same problem using insights from the discussion above and to shed light on the discrepancy. 

\citet{quataert_2004} considered three values for $\eta$: 0, 2 and 3. Both $\eta=0$ and $\eta=2$ pass through the Chandra measurement of the density at 10''. As for the other measurement at 2'', $\eta=2$ slightly overestimates the density, while $\eta=0$ slightly underestimates the density. Assuming uncertainty of a factor of 2, any value of $\eta$ between 0 and 2 is consistent with these data points.

The other set of observational data in that paper is the X-ray surface brightness at radii below 3.5''. In order to compare to the observational data, Quataert calculated the X-ray luminosity according to free-free emission, and then scaled the curve such that it will coincide with the measurement at 0.5''. When the comparison is performed in this way, $\eta=3$ seems closer to the data points than the other models. We note that all models greatly overestimate the data points at large radii. If, instead, the theoretical curves were calibrated to coincide with the observation at 3.5'', then all models would fit the data point equally well, up to a distance of 1'', which is Chandra's PSF. The discrepancy could be the result of an unresolved X - ray source close to SGR A*, which is not related to the gas. Another possible explanation for the discrepancy between the analytic model and the observational data is the centrifugal barrier, where the isotropic flow turns to an accretion disc.

In fact, as was discussed in section  \ref{sec:trunction}, close to the inner edge of the wind generating zone the solution is hydrostatic and density and temperature are constant, regardless of $\eta$. Therefore, it is no surprise that this measurement cannot distinguish between models with different values of $\eta$. Also, the observational data agrees with a flat density and temperature profiles of the gas. 

\subsection{Radio Signal from Tidal Disruption Events}

It was suggested that radio signals from tidal disruption events result from the interaction of outflow with the ambient gas \cite{giannios_metzger_2011}. This outflow might be a jet \citep{alexander_et_al_2016} or unbound debris \citep{krolic_et_al_2016}, and when it collides with the gas it creates a shock wave, which emits synchrotron radiation, like in supernova remnants \citep{chevalier_1998}. Tidal disruption events serve as a probe to the gas density at the centre of another galaxy.

Let us consider a shock wave travelling through the diffuse gas, and the synchrotron emission from such a shock. We make the standard assumptions, i.e. that a certain fraction of the shocked equipartition energy goes into generating the magnetic field and Fermi acceleration of particles to a power law distribution in energy $d n_e / d \gamma \propto \gamma^{-p}$, where $\gamma$ is the Lorentz factor and $p$ is a constant. This problem has been considered in previous works \citep{leventis_et_al_2012}, but we repeat here the main results. We will consider four different combinations 
of outflow velocities and optical depths. First, we consider whether the emitting region is optically thin or thick. Second, we consider whether the mass of the outflow is greater (i.e. matter moves at a constant velocity) or smaller (outflow decelerates) than the mass of the swept up gas. If the outflow moves at a constant velocity and is optically thick then $d \ln L / d \ln t = 2$. If the decelerating, optically thick shock wave moves into a medium with density $n \propto r^{-k}$, then $r \propto t^{2/\left(5-k\right)}$ and $d \ln L / d \ln t = \frac{2 \left( k-1\right)}{5-k}$. In the optically thin case with constant velocity,  $d \ln L / d \ln t = 3-\frac{k}{4} \left(p+5\right)$, while for the decelerating case $d \ln L / d \ln t = \frac{8 k - 4 p \left(k - 3\right) + 3 p - 21}{2 \left(k - 5\right) }$.

We note that the light-curve of an optically thick, constant velocity outflow is independent of the density slope. Also, the light-curve of an optically thin, decelerating outflow is weakly dependent on $k$: for $p = 2.5$, $d \ln L / d \ln t$ changes from $-1.7$ to $-2.1$ as $k$ varies from 0 to 2. For this reason, density breaks will always lead to chromatic (frequency dependent) breaks in the radio lightcurves.

If the density profile can be inferred from the radio signal of a tidal disruption event, it might be possible to infer from it the distribution of wind emitting stars. We recall that at radii larger than the stagnation radius the density scales as $n \propto r^{1-\eta}$. Hence, if it is possible to observe the radio signal to late enough times, and assuming that the distribution of the stars is not truncated, then the distribution of wind emitting stars can be calculated.

Interestingly, the exact analytic solution we have derived (for the special case of $\eta=5/2$) may apply reasonably well to the central parsecs of the rare E+A (or post-starburst) galaxies that are highly overrepresented among TDE hosts \citep{french_arcavi_zabludoff_2016}.  It is not yet understood why TDEs are observed with such high frequency in E+A galaxies, but one plausible explanation is that their recent starbursts produce unusually dense and steeply sloped nuclear star clusters, where the rate of two-body stellar scatterings is high and stars are frequently deflected into the supermassive black hole.  In order to explain the delay time distribution of observed TDEs, this hypothesis likely requires density profiles much steeper \citep{2017arXiv170900423S} than the usual Bahcall-Wolf \citep{bahcall_wolf_1977} equilibrium profile (i.e. $\eta > 7/4$), and resolved observations of one of the nearest E+As do find $\eta \approx 2.3$ on $\approx 4~{\rm pc}$ scales \citep{stone_van_velzen_2016}. 

\section{Summary} \label{sec:summary}
In this paper we studied the flow of stellar winds in an isotropic star cluster around a super massive black hole. We showed such a flow has a stagnation point close to the Bondi radius. All the material injected inside the stagnation radius will accrete onto the black hole, while material injected outside the stagnation radius will fly out to infinity. Depending on the parameters, there could be sonic points outside and/or inside the stagnation radius. Sonic points divide space into two regions such that matter and information can only travel from one domain to another. In our case, matter and information can only travel from the region containing the stagnation point. 

This system exhibits a rich collection of qualitatively different behaviours depending on its parameters (i.e. the adiabatic index of the gas and the slope of the stellar density profile). We obtained analytic expressions for the asymptotic behaviour in all cases that are relevant for astrophysics. In one particular case we were also able to obtain a completely analytic solution for the entire hydrodynamic profile. In all other cases we can obtain the profile by numerically integrating the ordinary differential equations. We verified these profiles with a time dependent hydrodynamic simulation. 

The biggest challenge in obtaining an analytic solution is determining the location of the stagnation point. We have shown that the condition is that the hydrodynamic profiles pass smoothly through the stagnation point. For arbitrary values of the stagnation radius it would still be possible to integrate the hydrodynamic profiles from both extremes ($r\ll R_b$ and $r\gg R_b$) to the stagnation point, but they would not connect smoothly at the stagnation point.

Having solved the hydrodynamic problem, we proceeded to discuss radiation. We stipulated the conditions for radiative cooling to be important. Namely, these conditions are that the fluid element spends enough time in a region for it to lose a considerable amount of energy, and that radiative cooling exceed wind heating. In cases where both conditions are satisfied then radiation becomes important and our analysis cannot
be applied.

Even if radiation does not change the dynamics of the problem, it is still important from an observational point of view. In some cases the entire system can be transparent, even if the wind profile extends indefinitely. We calculate the Bremsstrahlung spectrum in such cases and describe a practical manner by which an observation of the X-ray luminosity and spectral break frequency can be translated into an estimate of the mass inflow rate onto the central supermassive black hole (eq.~\ref{eq:Mdotcalib}). 

%If, on the other hand, the system is opaque, and photon production is fast enough then an outside observer will simply see a blackbody spectrum from the photosphere, perhaps with excess in low energy photons due to contribution from exterior shells.

The insights from this study helped us to resolve a difficulty pointed out by a previous work that dealt with our galactic centre \citep{quataert_2004}. In that work, the author solved the steady state hydrodynamic equations with a source term and gravity, and for different values of $\eta$. He showed that one set of observations was best fit by $\eta=0$, while another by $\eta=3$. We pointed out the error in the methodology, namely, choosing a problematic anchor in the experimental data. After the error is corrected all models agree with both observations. More observations are therefore necessary to constrain $\eta$.

Finally, we applied our model to the gas environment of other quiescent galactic nuclei. Information about these nuclei can be extracted from the radio signal of tidal disruption events. From these measurements it is possible to extract the density as a function of distance from the centre of a galaxy. Our models can, in principle, relate the inferred density profile to other properties of that galactic nucleus, such as mass of the black hole and the distribution of wind emitting stars. We note that other models attribute features in the radio signals to the outflow rather than the diffuse medium. Hopefully, future observations of tidal disruption events will provide data which will allow us to test these competing models.

\section*{Acknowledgements}

AY would like to thank Jeremiah Ostriker for the enlightening discussion.

%%%%%%%%%%%%%%%%%%%%%%%%%%%%%%%%%%%%%%%%%%%%%%%%%%

%%%%%%%%%%%%%%%%%%%% REFERENCES %%%%%%%%%%%%%%%%%%

% The best way to enter references is to use BibTeX:

\bibliographystyle{mnras}
\bibliography{references} % if your bibtex file is called example.bib

% Alternatively you could enter them by hand, like this:
% This method is tedious and prone to error if you have lots of references
%\begin{thebibliography}{99}
%\bibitem[\protect\citeauthoryear{Author}{2012}]{Author2012}
%Author A.~N., 2013, Journal of Improbable Astronomy, 1, 1
%\bibitem[\protect\citeauthoryear{Others}{2013}]{Others2013}
%Others S., 2012, Journal of Interesting Stuff, 17, 198
%\end{thebibliography}

%%%%%%%%%%%%%%%%%%%%%%%%%%%%%%%%%%%%%%%%%%%%%%%%%%

%%%%%%%%%%%%%%%%% APPENDICES %%%%%%%%%%%%%%%%%%%%%

%\appendix

%\section{Some extra material}

%If you want to present additional material which would interrupt the flow of the main paper,
%it can be placed in an Appendix which appears after the list of references.

%%%%%%%%%%%%%%%%%%%%%%%%%%%%%%%%%%%%%%%%%%%%%%%%%%

% Don't change these lines
\bsp	% typesetting comment
\label{lastpage}
\end{document}